\documentclass[useAMS,usenatbib]{mn2e} 
\usepackage[dvips]{graphicx}
\usepackage{wasysym} 
\usepackage{threeparttable}

\newcommand{\nh}{$\rm{N}_{\rm H}$} 
\newcommand{\av}{$\rm{A}_{\rm V}$}

\title{The Relation Between Optical Extinction and Hydrogen Column
  Density in the Galaxy}

\author[Tolga G\"uver and Feryal \"Ozel]{Tolga G\"uver and Feryal
\"Ozel \\ University of Arizona, Department of Astronomy, 933
N. Cherry Ave., Tucson, AZ, 85721}

\begin{document}
\date{}
\pagerange{\pageref{firstpage}--\pageref{lastpage}} \pubyear{2009}
\maketitle

\label{firstpage}

\begin{abstract} A linear relation between the hydrogen column density
  (\nh) and optical extinction (\av) in the Galaxy has long been
  observed. A number of studies found differing results in the slope
  of this relation. Here, we utilize the data on 22 supernova remnants
  that have been observed with the latest generation X-ray
  observatories and for which optical extinction and/or reddening
  measurements have been performed and find \nh $(cm^{-2}) = (2.21\pm
  0.09)\times 10^{21}\,$ \av (mag).  We compare our result with the
  previous studies and assess any systematic uncertainties that may
  affect these results.
\end{abstract}


\section{Introduction}

Photoelectric absorption by interstellar matter causes rapid
attenuation of the observed soft X-ray spectra in all Galactic
sources.  Measuring the amount of the X-ray extinction yields
information on the total column density of the interstellar matter
between the observer and the source, because any element that is not
fully ionized contributes to the extinction of X-rays.  Although the
amount of the X-ray extinction is generally expressed in terms of the
equivalent hydrogen column density (\nh), it is generally caused,
especially above $\sim$0.25~keV, by the most abundant heavier elements
like O, Ne, Fe, Mg, and Si.

Optical extinction, on the other hand, is caused by grains composed of
these same heavier elements. Therefore, the relation between optical
and X-ray extinction depends on the relative abundances of metals and
the temperature of the intervening medium, which primarily determine
the relative fraction of metals in grains versus in the gas state. For
sources, however, that lie at large distances from the observer,
photons traverse many different regions of the interstellar medium
(ISM), and, thus, sample a wide variety of physical conditions. Under
the assumption that different lines of sight sample the same
distribution of physical conditions in the ISM, one expects a relation
between the optical extinction and the hydrogen column density towards
distant sources.

The relation between optical extinction and hydrogen column density
has been observationally studied for decades using various techniques
\citep[see, e.g.,][]{Reina:1973ph, Gorenstein:1975yj, Predehl:1995nu}.
Using measurements for two X-ray binaries together with the two
extended sources GCX and Cas A (SNR G111.7$-$02.1),
\cite{Reina:1973ph} derived the first relation \nh (cm$^{-2}$) =
1.85$\times$ 10$^{21}$\, \av (mag). \cite{Gorenstein:1975yj} found a
similar relation shortly afterwards using independent optical
extinction and column density measurements for 7 supernova remnants
(SNR), which yielded \nh (cm$^{-2}$) = (2.22 $\pm$0.14)$\times$
10$^{21}$\, \av (mag). Later, using ROSAT observations of 25 bright
X-ray point sources as well as of 4 SNRs, \cite{Predehl:1995nu}
determined the shape and intensity of the soft X-ray halos around
these sources, which they used in conjuction with dust halo models to
measure the dust column density. In this study, they also found a
relation between the X-ray derived hydrogen column density and optical
extinction given by \nh $ (cm^{-2}) = (1.79\pm 0.03) \times 10^{21}\,$
\av.

The three analyses quoted above resulted in relations that disagree
within their statistical uncertainties. This can be a result of {\it
  (i)} the low quality of early X-ray spectroscopic data used; {\it
  (ii)} the systematic uncertainties between the methods employed,
{\it (iii)} formal uncertainties because of the small number of
sources used, or {\it (iv)} strong variations in the ISM along the
various lines of sight. In this paper, we take advantage of the most
recent X-ray and optical observations of SNRs with high quality
instruments and compile the most complete set of extinction
measurements towards these sources to redetermine the relation between
the optical extinction and the hydrogen column density. We choose SNRs
because they do not suffer from significant intrinsic absorption that
may contribute to the measured extinction. In addition, owing to the
large number of observed sources and the high quality data from the
latest generation of X-ray detectors, we will be able to address which
of the above factors contribute to the variations in the \av $-$ \nh
~relation found in the previous studies.

In Section~2 we summarize the methods used for the measurement of the
hydrogen column density and optical extinction and present results
towards 21 supernova remnants. In Section~3, we report the resulting
\av $-$ \nh relation and discuss any systematic uncertainties that may
affect this result.

\begin{table*}
\centering
\caption{Optical extinction
  and X-ray measured hydrogen column densities as collected from
  literature. }
\begin{tabular}{ccccccccc}
  \hline\hline
  SNR Name & N$_{\rm H}$ & Model$^1$ & Ref. & E(B$-$V) & A$_{\rm V}$ & Method & Ref. \\
  & (10$^{22}$ cm$^{-2}$) & & & mag. & mag. & & \\
  \hline\hline
  SNR G0.0$+$0.0     & 14.7$\pm$1.2      & TP     &(1)  & 9.355$\pm$0.645 & 29$\pm$2       & NS                       & (2) \\
  SNR G004.5$+$06.8  & 0.52              & SE     &(3)  & 0.806$\pm$0.29  & 2.5$\pm$0.9    & FeII Ratio               & (4) \\
  SNR G6.5$-$0.1     & 0.47$\pm$0.1      & TP     &(5)  & 1.15$\pm$0.15   & 3.57$\pm$0.47  & H$_{\alpha}$/H$_{\beta}$ & (6) \\
  SNR G13.3$-$1.3    & 0.55$\pm$0.45     & TP     &(7)  & 0.15            & 0.47           & H$_{\alpha}$/H$_{\beta}$ & (7) \\
  SNR G53.6$-$2.2    & 0.78$\pm$0.4      & TP     &(8)  & 1.15$\pm$0.15   & 3.57$\pm$0.47  & H$_{\alpha}$/H$_{\beta}$ & (6) \\
  SNR G54.1$+$0.3    & 1.6$\pm$0.1       & PL     &(9)  & 2.581$\pm$0.226 & 8.0$\pm$0.70   & NS                       & (10) \\
  SNR G67.7$+$1.8    & 0.515$\pm$0.195   & TP     &(11) & 1.7$\pm$0.3     & 5.27$\pm$0.93  & H$_{\alpha}$/H$_{\beta}$ & (12) \\
  SNR G69.0$+$2.7    & 0.3$\pm$0.1       & PL     &(13) & 0.8             & 2.48           & H$_{\alpha}$/H$_{\beta}$ & (14) \\
  SNR G74.0$-$8.5    & 0.023$\pm$0.01    & TP     &(15) & 0.08            & 0.25           & NS                       & (16) \\
  SNR G109.1$-$1.0   & 1.12$\pm$0.3      & AE     &(17) & 0.995$\pm$0.205 & 3.15$\pm$0.65  & H$_{\alpha}$/H$_{\beta}$ & (18) \\
  SNR G111.7$-$02.1  & 1.25$\pm$0.03     & BB     &(19) & 1.613$\pm$0.129 & 5.0$\pm$0.40   & SII ratio                & (20) \\
  SNR G116.9$+$0.2   & 0.79$\pm$0.12     & TP     &(21) & 0.871$\pm$0.161 & 2.70$\pm$0.5   & H$_{\alpha}$/H$_{\beta}$ & (22) \\
  SNR G119.5$+$10.2  & 0.28$\pm$0.05     & TP+PL  &(23) & 0.410$\pm$0.132 & 1.27$\pm$0.41  & Extinction map           & (24) \\
  SNR G120.1$+$1.4   & 0.435$\pm$0.045   & TP     &(25) & 0.6$\pm$0.039   & 1.86$\pm$0.12  & NS                       & (26) \\
  SNR G130.7$+$3.1   & 0.416$\pm$0.08    & TP+PL  &(27) & 0.68            & 2.11           & H$_{\alpha}$/H$_{\beta}$ & (28) \\
  SNR G132.7$+$1.3   & 0.43$\pm$0.25     & TP     &(5)  & 0.710$\pm$0.040 & 2.201$\pm$0.124& H$_{\alpha}$/H$_{\beta}$ & (29) \\
  SNR G184.6$-$5.8   & 0.36$\pm$0.004    & PL     &(30) & 0.50$\pm$0.060  & 1.55$\pm$0.186 & Ly$\alpha$ Absorption    & (31) \\
  SNR G260.4$-$3.4   & 0.454$\pm$0.049   & 2BB    &(32) & 0.839           & 2.60           & NS                       & (33) \\
  SNR G263.9$-$3.3   & 0.0259$\pm$0.0001 & 2BB    &(34) & 0.074           & 0.38           & H$_{\alpha}$/H$_{\beta}$ & (35) \\
  SNR G296.5$+$10.0  & 0.1$\pm$0.01      & 2BB    &(36) & 0.161           & 0.50           & H$_{\alpha}$/H$_{\beta}$ & (37) \\
  SNR G327.6$+$14.6  & 0.07$\pm$0.01     & SI     &(38) & 0.11            & 0.34           & HI/GC                    & (39) \\
  SNR G332.4$-$00.4  & 0.7$\pm$0.2       & 2BB    &(40) & 1.516$\pm$0.29  & 4.70$\pm$0.90  & FeII Ratio               & (4)  \\
  \hline
\end{tabular}
\newline
\footnotesize{ \begin{flushleft} $^1$ Model abbreviations: TP: Thermal
    Plasma, PL:  Power-Law, BB:  Blackbody, AE: Absorption  Edge modeling,
    NS: Nearby Stars, SE: Synchrotron Emission, SI: Shocked ISM
  \end{flushleft} 
  References : (1)  \cite{Sakano:2004sr}; (2) \cite{Predehl:1994ni}; (3)
  \cite{Reynolds:2007pt};       (4)       \cite{Oliva:1989tt};       (5)
  \cite{Rho:1998ug};  (6) \cite{Long:1991sa};  (7) \cite{Seward:1995zn};
  (8) \cite{Saken:1995or}; (9) \cite{Lu:2002jf}; (10) \cite{Koo:2008fs};
  (11) \cite{Hui:2008vl};    (12)    \cite{Mavromatakis:2001ls};    (13)
  \cite{Li:2005la};        (14)        \cite{Hester:1989sj};        (15)
  \cite{Katsuda:2008sf};       (16)      \cite{Parker:1967sh};      (17)
  \cite{Durant:2006kc};       (18)       \cite{Fesen:1995jy};       (19)
  \cite{Hwang:2004cg};       (20)       \cite{Hurford:1996sd};      (21)
  \cite{Craig:1997am};        (22)       \cite{Fesen:1997bh};       (23)
  \cite{Slane:1997ft};     (24)     \cite{Mavromatakis:2000yl};     (25)
  \cite{Warren:2005ay};     (26)    \cite{Ruiz-Lapuente:2004mw};    (27)
  \cite{Gotthelf:2007ap};       (28)      \cite{Fesen:1988oe};      (29)
  \cite{Fesen:1995fu};       (30)       \cite{Massaro:2006wl};      (31)
  \cite{Sollerman:2000ch};       (32)       \cite{Hui:2006mb};      (33)
  \cite{Gorenstein:1975yj};     (34)     \cite{Manzali:2007zn};     (35)
  \cite{Manchester:1978dn};     (36)     \cite{De-Luca:2004oc};     (37)
  \cite{Ruiz:1983fd};        (38)        \cite{Acero:2007zi};       (39)
  \cite{Raymond:1995aj}; (40) \cite{De-Luca:2006ts}.
  \label{obs_data}}
\end{table*}

\section{Extinction and X-ray absorption measurements}

Of the 243 supernova remnants included in the extensive catalog of
\cite{Guseinov:2003jf, Guseinov:2004km, Guseinov:2004qw},
approximately 143 sources have been observed with the {\it Chandra},
XMM-{\it Newton}, or {\it Suzaku} satellites.  The high quality
spectral data obtained with these satellites allowed for a more
precise measurement of the hydrogen column density by modeling both
the X-ray continuum and the line features in the source spectra. We
compiled the hydrogen column density measurements for these 143
sources (see Table~1 for the relevant references).

We then searched for independent measurements of the optical
extinction towards all the SNRs listed in the \cite{Guseinov:2003jf,
  Guseinov:2004km, Guseinov:2004qw} catalog.  Unlike the case of X-ray
extinction, our search revealed that there is a lack of reported
measurements of optical extinction or reddening: Out of the 243 SNRs,
we were able to obtain only 22 independent $A_{\rm V}$ measurements.
Luckily, this set of 22 sources is a subset of the 143 sources, for
which we have high-resolution measurements of the X-ray extinction
with the new X-ray satellites.

We discuss here the data as well as the methods that are used to
determine the hydrogen column density and optical extinction. This
will be important in quantifying systematic uncertainties in the
correlation between optical and X-ray extinction.

\subsection{X-ray Hydrogen Column Density Measurements}

Hydrogen column density measurements of supernova remnants in the
X-rays are performed by modeling their spectra, usually in the
$0.2-8.0$~keV range. Extinction is determined by comparing an
intrinsic source spectrum, such as a blackbody, a power-law, or a
thermal plasma emitting bremsstrahlung radiation, with the observed
spectrum. The model parameters inferred from spectral fits are often
correlated with the hydrogen column density, especially when the
analyzed data cover a small range in the X-ray band and the energy
resolution is low \citep[see, e.g., Figure 2 of][]{Predehl:1995nu}.
Moreover, when the spectra of extended sources are observed with
detectors of low angular resolution, the intrinsic variations of the
source spectra that cannot be resolved bias the results.

Observations with the Chandra, XMM-Newton, and Suzaku satellites
\citep[see, e.g.,][]{Paerels:2003mz, Juett:2004jy, Juett:2006bl} yield
significant improvements in both of these areas. The increased angular
resolution of the telescopes make it possible to model small regions
within the SNR and detect local changes in the intrinsic spectral
properties.  At the same time, the increased energy resolution and
signal to noise ratios lead to better modeling of both the continuum
spectra and of line features. In the most favorable cases, e.g., SNR
G109.1$-$1.0, X-ray absorption edges of heavier elements, such as Mg
and Ne, can be directly detected, completely eliminating the
dependence of the column density on the assumed intrinsic X-ray
spectrum of the source \citep{Durant:2006kc}. All of these lead to
better constrained hydrogen column densities compared to earlier
studies.

In the majority of the SNRs in our study, the X-ray spectra were fit
with a thermal plasma model, where emission lines were also taken into
account whenever possible (see Table~1). The remaining spectra were
modeled either by a power-law or a combination of two blackbodies. In
order to make the analysis as free from the assumed continuum model as
possible, we did not exclude from the study the reported hydrogen
column density measurements where other continuum models were used.

For the spatially resolved SNRs, hydrogen column densities measured
towards different regions of the remnant can show variations that are
larger than the statistical errors of each measurement. In these
cases, we adopted the average of these values with an error that
accounts for the observed scatter. Finally, in the cases where the
central point sources can be resolved, we also used the \nh
measurement obtained by modeling the emission from the central neutron
star surface either by a blackbody, a power-law, or a hydrogen
atmosphere model.

\subsection{Optical Extinction Measurements}

Determining the optical extinction towards SNRs is more challenging
than the hydrogen column density measurements in the X-rays and
demands high signal to noise ratio spectra. Nearly all of the methods
involve measuring the reddening using the intensity ratio of two
emission lines and converting the reddening into an optical
extinction.

One of the well known and reliable methods of measuring the extinction
is based on the Balmer decrement, which involves the intensity ratio
of the observed H$_{\alpha}$ (6563 \AA) and H$_{\beta}$ (4861 \AA)
emission lines.  The observed relative intensity is compared to that
expected for a gaseous nebula with typical temperature and electron
density.  Since this ratio depends very weakly on the physical
conditions of the nebula, the theoretical ratio can be calculated with
minimal uncertainty. Thus, the comparison allows a measurement of the
reddening, and hence, of the optical extinction \citep[see,
e.g.,][]{Osterbrock:1989xe, Lequeux:2005sf}.  The majority of the
optical extinction measurements presented here are found using this
method since both of the lines lie in the optical range and, in most
cases, are strong enough to be resolved.

One other frequently used method involves measuring the SII multiplet
ratios \citep{Miller:1968fu} in the infrared ($\sim$10320 \AA) and
blue ($\sim$4068\AA). As with the Balmer decrement method, this ratio
also depends only very weakly on the temperature and density in the
remnant. However, the disadvantage of this method is the fact that it
is not always practical to perform spectral observations in both of
these wavelength regions with high enough spectral resolution.

In a third method, the two most prominent IR transitions of Fe[II]
(1.6435 $\mu$m and 1.2567$\mu$m) that arise from the same upper level
can be used. Their intensity ratio does not depend strongly on the
temperature and density of the gas \citep{Oliva:1989tt}, allowing for
a determination of the extinction.  In our sample, the optical
extinction towards SNR G332.4-00.4 and SNR G004.5+06.8 has been
measured with this method by \citep{Oliva:1989tt}, using a theoretical
ratio of 1.36 for these lines \citep{Nussbaumer:1988xe}.

A final method involves using nearby stars with known distances to
estimate the extinction towards the SNR. In Table~1, we denoted the
four optical extinction measurements obtained with this method as
``nearby stars''.

In Table~\ref{obs_data} we present the compilation of the optical
extinction data. For the cases where only the reddening E(B$-$V)
measurement was reported, we converted these to an extinction (\av)
assuming a $A_{\rm V} = 3.1 E(B-V)$ relation \citep[see,
e.g.,][]{Fitzpatrick:2004rr, Savage:1979ly}.  Finally, in the absence
of any reported systematic or statistical errors, we assigned a 15\%
uncertainty on the measurements for the purposes of the fit only,
which is similar to the average errors given for the other
measurements.  These cases are left without errors in Table~1.

\begin{figure*}
\centering    \includegraphics[scale=0.45]{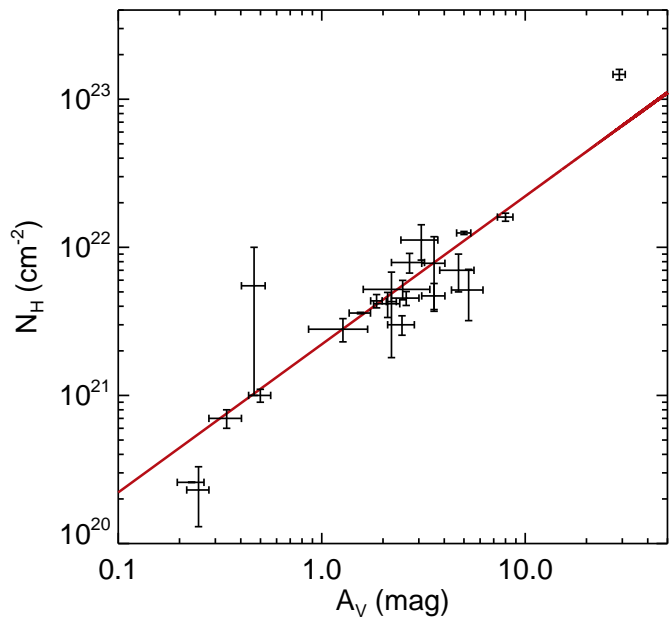}    
\caption{The observed correlation between hydrogen column density and
  optical extinction, together with the best fit linear model, found
  as \nh $= (2.21 \pm 0.09) \times 10^{21} \times$\av.}
\end{figure*}

\section{Results and Discussion}

We plot in Figure~1 the optical extinction, \av, for 22 supernova
remnants against their hydrogen column density, measured from their
high energy resolution X-ray spectra. Figure~1 includes a typical 15\%
error on the optical extinction for those remnants that did not have
reported errors on this measurement (see Table~1).  We performed a
linear fit between these quantities and obtained the best-fit relation

\begin{eqnarray}
\rm{N}_{\rm H}(cm^{-2}) = (2.21\pm 0.09)\times 10^{21}\, \rm{A}_{\rm V},
\end{eqnarray}

where the errors correspond to 1$-\sigma$ statistical uncertainty.  We
present the best-fit line in Figure~1.  This relation can also be
expressed as \nh $(cm^{-2}) = (6.86 \pm0.27)\times 10^{21}\, E(B-V)
(mag)$ between the reddening and the hydrogen column density.

We then investigated whether the continuum models that are used to fit
the X-ray observations to measure the hydrogen column density affect
the results found for the \av $-$ \nh relation. Specifically, we
looked into the power-law continuum models, which are known to give
higher values for the hydrogen column density than others.  To this
end, we compared the \av $-$ \nh ~relation that we found from fitting
the entire sample to that derived when the three power-law continuum
fits in the sample were excluded. We found that the resulting
relations do not differ significantly from each other either in the
slope or in the errors.  Specifically, excluding the subset of
observations where the X-ray continuum was modeled with a power-law
results in the relation \nh $= (2.29 \pm0.11) \times10^{21} \times$
\av, while all the data yield \nh $ = (2.21 \pm0.09) \times 10^{21}
\times$ \av.

Another possible systematic uncertainty is introduced by the use of
solar abundances when measuring the hydrogen column density from X-ray
spectra.  It has been shown that different assumptions on the
elemental abundances (e.g., solar vs. ISM) in the interstellar medium
can lead to $\sim 5\%$ variation in the hydrogen column density
\citep{Wilms:2000oy}.  The \nh values used in our study could be
subject to similar systematic uncertainties.

Comparing our result with previous studies, we find that it is in very
good agreement with the relation given by \cite{Gorenstein:1975yj} \nh
$(cm^{-2}) = (2.22 \pm 0.14) \times 10^{21}\,$ \av, while it differs
at the $3\sigma$ level from the result obtained by
\cite{Predehl:1995nu} \nh $(cm^{-2}) = (1.79 \pm 0.03) \times
10^{21}\,$ \av.  This difference can be attributed to a number of
causes: ROSAT's narrow bandpass, which does not allow tight
constraints on the intrinsic spectral shape; the authors' assumption
that the intrinsic spectra of all their sources are power laws;
possible intrinsic absorption; and the uncertain properties of the
optical counterparts, which affect the assumed extinction values.

Despite the improvements on the number and quality of the
observations, there may still be a number of systematic errors and
selection effects present in our study. A larger, homogeneous sample
of supernova remnants, studied with both the last generation X-ray
satellites as well as with sensitive optical spectrographs will help
refine the studies of the interstellar medium and the relation between
the optical extinction and the hydrogen column density in the Galaxy,
hence the dust to gas ratio. Increasing the number of sources in the
sample may also allow for a search for variations in the relation
towards different lines of sight.

\section*{Acknowledgments}
We would like to thank the anonymous referee for very useful comments.
This work was supported by NSF grant AST 07-08640.

\bibliography{bibtex_file} 
\bibliographystyle{mn2e}

\label{lastpage}

\end{document}